\documentstyle[preprint,prc,aps,amsmath]{revtex}
\tightenlines

\begin{document}

\draft
\title{$\eta$ meson production in $NN$ collisions}
\author{K. Nakayama$^{a,b}$, J. Speth$^b$ and T.-S. H. Lee$^c$}
\address{$^a$Department of Physics and Astronomy, University of Georgia, 
Athens, GA 30602, USA \\
$^b$Institut f\"ur Kernphysik, Forschungszentrum-J\"ulich,
D-52425, J\"ulich, Germany \\
$^c$Physics Division, Argonne National Laboratory, Argonne, IL 60439, USA}
\maketitle

\begin{abstract}
$\eta$ meson production in both proton-proton and proton-neutron collisions 
is investigated within a relativistic meson exchange model of hadronic 
interactions. It is found that the available cross section data can be described
equally well by either the vector or pseudoscalar meson exchange mechanism for 
exciting  the $S_{11}(1535)$ resonance. It is shown that the 
analyzing power data can potentially be very useful in distinguishing 
these two scenarios for the excitaion
of the $S_{11}(1535)$ resonance. 
\end{abstract}

\vskip 0.5cm


\newpage

\section{ Introduction}

The production of $\eta$ mesons in nucleon-nucleon ($NN$) collisions near the threshold 
energy has been a subject 
of considerable interest in the past few years, since the existing data are by far
the most accurate and complete among those for heavy meson production. Consequently, 
they provide an opportunity to investigate this reaction in much more detail than any of the other 
heavy meson production reactions. In addition to the total cross section for the $pp \rightarrow 
pp\eta$ reaction \cite{spes3eta1,spes3eta2,celsiuseta,cosy11eta,Calenp,Tatischeff}, we now have data 
for $pn \rightarrow pn\eta$ \cite{Calenn} and $pn \rightarrow d\eta$ \cite{Calenp,Calend} reactions. 
The differential cross section data for the $pp \rightarrow pp\eta$ reaction 
\cite{Calenh} are also available. Consequently, there have been a large number of theoretical 
investigations on these reactions \cite{THEORY,Bernard,Batinic,Wilkin,Gedalin,Santra,Pena}. 

The production of $\eta$ mesons in 
$NN$ collisions is thought to occur predominantly through the excitation (and de-excitation) of 
the $S_{11}(1535)$ resonance, to which the $\eta$ meson couples strongly. However, the excitation 
mechanism of this resonance is currently an open issue. For example, Batini\'c et al.
\cite{Batinic} have found that 
both $\pi$ and $\eta$ exchanges are the dominant excitation
mechanisms. 
However, they have considered only the $pp \rightarrow pp\eta$ reaction. F\"aldt and Wilkin 
\cite{Wilkin} and Gedalin et al. \cite{Gedalin} have considered both the $pp \rightarrow pp\eta$ 
and $pn \rightarrow pn\eta$ reactions. In the analysis of Ref.\cite{Wilkin} the 
$pn \rightarrow d\eta$ reaction is also considered. These authors \cite{Wilkin,Gedalin} find  
$\rho$ exchange to be the dominant excitation mechanism of the $S_{11}(1535)$ resonance. 
In particular, it has been claimed \cite{Wilkin} that $\rho$ meson exchange is important for 
explaining the observed shape of the angular distribution of the $pp \rightarrow pp\eta$ 
reaction. In an anaylsis of $pp \rightarrow pp\eta$ reaction, Santra and Jain \cite{Santra}
also considered $\rho$ meson exchange as the dominant excitation mechanism of the 
$S_{11}(1535)$ resonance. In contrast to the 
findings of Refs.\cite{Batinic,Wilkin,Gedalin,Santra},
 Pe\~na et al. \cite{Pena} have  found that the dominant 
contribution arises, not from the $S_{11}(1535)$ resonance current,
 but from the shorter range part of the nucleonic currents.
In this work,  we shall report on another possible scenario for 
exciting the $S_{11}(1535)$ resonance that reproduces both the $pp \rightarrow pp\eta$ and 
$pn \rightarrow pn\eta$ reactions and discuss the possibility of disentangling these reaction 
mechanisms. 

Although we focus here on the problem just mentioned, the description of 
$\eta$ meson production in $NN$ collisions presents other interesting aspects. For example, 
the $\eta$ meson interacts much more strongly with the nucleon than do mesons like the pion 
so that not only the $NN$ final state interaction (FSI), but also the $\eta N$ FSI is likely 
to play an important role, 
thereby offering an excellent opportunity to learn about the $\eta N$ interaction 
at low energies. In fact, the near-threshold energy dependence of the observed total 
cross section for $\eta$ meson production differs from that of $\pi$ and $\eta^\prime$ production, 
which follow the energy dependence given simply by the available phase-space together with the $NN$ 
FSI. The enhancement of the measured cross section at small excess energies in $\eta$ production 
compared to those in $\pi$ and $\eta^\prime$ production is generally attributed to the 
strong attractive $\eta N$ FSI. 
In addition to all of these issues, the theoretical understanding of $\eta$ meson production in $NN$ 
collisions near threshold in free space is also required for investigating the dynamics of the 
$S_{11}(1535)$ resonance in the nuclear medium, the possible existence of $\eta NN$ bound states, and 
the possibility of using $\eta$ to reveal the properties of high-density nuclear matter created 
in relativistic heavy-ion collisions.  

In section II we introduce our model and define the meson
 production currents whose details are given in the appendix. An alternative model, which
is similar to previous works based on $\rho$ exchange dominance,
 is introduced in section III. The results are given in section IV.
Section V provides a summary.

\section{The Meson Exchange Model}
\begin{figure}
\vspace{15cm}
\includegraphics{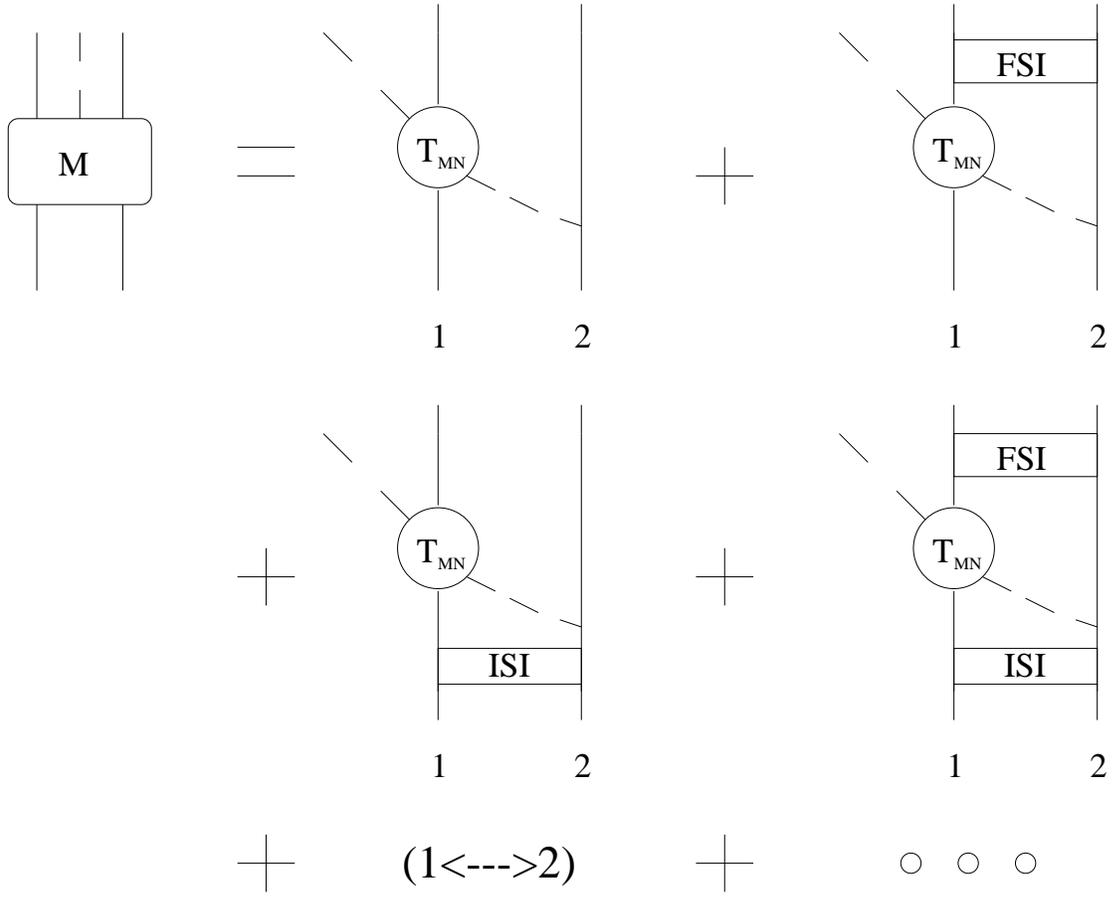}
\caption{Amplitude for the $NN\rightarrow NN\eta$ reaction considered in the present work.
$T_{MN}$ denotes the $MN$ $T$-matrix. ISI and FSI stand for the initial and final state 
$NN$ interaction, respectively.}
\label{fig0}
\end{figure}

Our model of the $NN\rightarrow NN\eta$ reaction
is based on a relativistic meson exchange model of hadronic 
interactions. The  reaction amplitude is calculated in
the Distorted Wave Born 
Approximation. Here we follow a (non-rigorous but otherwise economic) diagrammatic approach 
to present our formulation . 
A more rigorous derivation of the reaction amplitude will be reported 
elsewhere. We start by considering the meson-nucleon ($MN$) and $NN$ interactions 
as the building blocks for constructing the total amplitude describing the $NN\rightarrow
NNM$ reaction. We then consider all possible combinations of these building blocks in a 
topologically distinct way, with two nucleons in the initial state and two nucleons plus a 
meson in the final state. In this process of constructing the total amplitude, care must 
be taken in order to avoid diagrams that lead to double counting. Specifically,  
the diagrams that lead to mass and vertex renormalizations must
be excluded since we choose to use the physical masses and coupling constants.
The resulting amplitude constructed in this 
way is displayed in Fig.~\ref{fig0}. The ellipsis indicates those diagrams that are more 
involved numerically (including, in particular, the $MN$ FSI, which otherwise would be 
generated by solving the three-body Faddeev equation). So far there are very few attempts 
to account for them \cite{Gedalin,Moalem}. 
These higher order terms are also not considered in this work.

In order to make use of the available potential models 
of $NN$ scattering, we will carry out our 
calculation within a
three-dimensional formulation which is 
deduced from Bethe-Salpeter formulation by restricting the
propagating two nucleons to be on their mass shell. We follow the
procedure of Blankenbecler and Sugar\cite{blank}. The meson production 
amplitude illustrated in Fig.1 then takes the
following familar form
\begin{equation}
M = (1 + T^{(-)\dagger}_f i G^{(-)*}_f) J (1 + i G^{(+)}_i T^{(+)}_i) \ ,
\label{ampl0}
\end{equation}
where $T_{(i,f)}$ denotes the $NN$ $T$-matrix interaction in the initial($i$)/final($f$) 
state, and 
$G_{(i,f)}$ is the three-dimensional
Blankenbecler-Sugar (BBS) propagator.
The superscript $\pm$ in $T_{(i,f)}$ as well as in $G_{(i,f)}$ in Eq.(\ref{ampl0}) 
indicates the boundary condition, $(-)$ for incoming and 
$(+)$ for outgoing waves.
 The production current is denoted by $J$, which is defined by 
the $MN$ $T$-matrix with one of the meson legs attached to a nucleon (first diagram
on the r.h.s. in Fig.~\ref{fig0}),
\begin{equation}
J = \sum_{M'} \left[T_{(MN\leftarrow M'N)}\right]_1iP_{M'}\left[\Gamma_{M'NN}\right]_2
  +  (1 \leftrightarrow 2) \ ,
\label{curr0}
\end{equation}
where $T_{(MN\leftarrow M'N)}$ stands for the $MN$ $T$-matrix describing the transition 
$M'N \rightarrow MN$, $\Gamma_{M'NN}$ and $P_{M'}$ stand for the $M'NN$ vertex and the 
corresponding meson propagator respectively. 
The subscripts 1 and 2 stand for the two interacting 
nucleons 1 and 2. The summation runs over the intermediate meson $M'$.
Eq.(\ref{ampl0}) is the basic 
formula on which the present calculation is based. 
We note\cite{notelee} here that care has been taken in the above
three-dimensional formulation to avoid double counting problems.

In the near threhold energy region, the two nucleon energy in the
final state $f$ is very low and hence the $NN$ FSI amplitude, $T^{(-)\dagger}_f$ in Eq.(1),
can be calculated from 
a number of realistic $NN$ potential models in the literature.
In the present work 
we use the $NN$ model developed by the Bonn group\cite{MHE87}
to calculate the FSI. 
This model is defined by 
 a three dimensionally 
 reduced BBS version of 
the Bethe-Salpeter equation 
\begin{equation}
T = V + ViGT \ ,
\label{scateqn}
\end{equation}
where $G$ denotes the BBS two-nucleon propagator, consistent with those appearing 
in Eq.(\ref{ampl0}). (note that our definition of $V (T)$ 
differs from that of Ref.\cite{MHE87} by a factor of $-i$).

The $NN$ initial state interaction (ISI) amplitude, $T^{(+)}_i$ 
 in Eq.(\ref{ampl0}), must be calculated
at incident beam energies above $1.25\ GeV$.
There exists no accurate $NN$ model for performing such a calculation.
For example, the model developed in
Ref.\cite{Lee} can only give a very qualitative description of the
$NN$ scattering phase shifts at energies above 1 GeV. 
In the present work we therefore follow 
Ref.\cite{Hanhart} and make the on-shell approximation to evaluate the ISI contribution.
This amounts to keeping only the $\delta-$function part of the
Green function $G_i$ in evaluating the loop integration
involving $i G^{(+)}_i T^{(+)}_i$. 
The required on-shell $NN$ ISI amplitude is obtained from Ref.\cite{SAID}.
As has been discussed in 
Ref.\cite{Hanhart}, this is a reasonable approximation to the full $NN$ ISI. In this approximation, 
the basic effect of the $NN$ ISI is to reduce the magnitude of
the meson production cross section. In fact, it is
easy to see that the angle-integrated production cross section in each
partial wave state $i$ is reduced by a factor of 
\cite{Hanhart} 
\begin{eqnarray}
\nonumber
\lambda_i & = & \left| \frac{1}{2} \left(\eta_i(p) e^{i2\delta_i(p)} + 1 \right) \right|^2
 \\     & = & \eta_{i}(p) \cos ^2(\delta_{i}(p)) + 
              \frac{1}{4}[1-\eta_{i}(p)]^2 \leq \frac{1}{4}[1+\eta_{i}(p)]^2 \ .
\label{isieff}
\end{eqnarray}
In the above equation, $\delta_i(p)$ and $\eta_i(p)$
denote the phase shift and corresponding inelasticity, respectively; $p$ stands for the relative
momentum of the two nucleons in the initial state.

There are a number of different approaches in the literature which model the production current
$J$ defined in Eq.(\ref{curr0}) based on meson exchange models. Following Refs.\cite{Nak2,Nak3},
we split the $MN$ $T$-matrix of Fig.1 
into the pole ($T^P_{MN}$) and non-pole ($T^{NP}_{MN}$) parts and 
calculate 
the non-pole part in the Born approximation. 
Then, the $MN$ $T$-matrix can be written as\cite{Pearce}
\begin{equation}
T_{MN} = T^P_{MN} + T^{NP}_{MN} \ ,
\label{MNT}
\end{equation}
where
\begin{equation}
T^P_{MN} = \sum_B f_{MNB}^\dagger i g_B f_{MNB} \ ,
\label{MNTP}
\end{equation}
with $f_{MNB}$ and $g_B$ denoting the dressed
 meson-nucleon-baryon ($MNB$) vertex and baryon
propagator, respectively. The summation runs over the relevant baryons $B$. The non-pole part
of the $T$-matrix is given by
\begin{equation}
T^{NP}_{MN} = V^{NP}_{MN} + V^{NP}_{MN}iGT^{NP}_{MN} \ ,
\label{MNTNP}
\end{equation}
where $V^{NP}_{MN}\equiv V_{MN} - V^P_{MN}$, with $V^P_{MN}$ denoting the pole part of the 
full $MN$ potential $V_{MN}$.  
$V^P_{MN}$ is given by equation analogous to Eq.(\ref{MNTP}) with the dressed vertices 
and propagators replaced by the corresponding bare vertices and 
propagators. We neglect the second term of Eq.(7) and hence the
full $MN$ $T$-matrix in Eq.(\ref{curr0}) is approximated 
as $T_{MN} \cong T^P_{MN} + V^{NP}_{MN}$.

\begin{figure}
\vspace{15cm}
\includegraphics{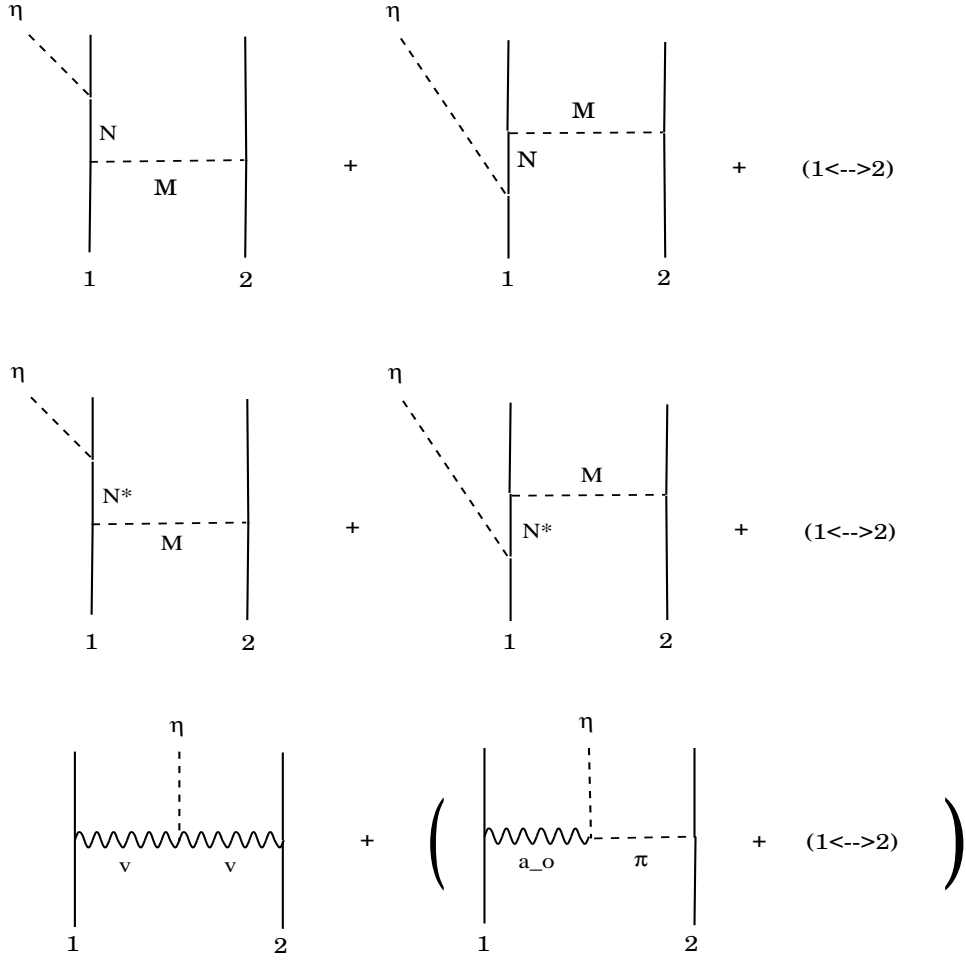}
\caption{$\eta$ meson production currents included in 
the present study. Upper row: nucleonic current $J_{nuc}$,  
$M = \pi, \eta, \rho, \omega, \sigma, a_o$.
Middle row: nucleon resonance current $J_{res}$, $N^*=S_{11}(1535),~P_{11}(1440)$ 
and $D_{13}(1520)$, $M = \pi, \eta, \rho, \omega$.  
Lower row: mesonic current $J_{mes}$, $v = \rho, \omega$.}
\label{fig1}
\end{figure}

With the approximation described above, the resulting
current $J$ consists of baryonic and mesonic currents. 
The baryonic current is further divided into 
the nucleonic and nucleon resonance ($N^*$) currents, so that the total current is written as
\begin{equation}
J = J_{nuc} + J_{res} + J_{mec} \ .
\label{curr}
\end{equation}
The individual currents are illustrated diagrammatically in Fig.~\ref{fig1}. 
Note that they are all Feynman diagrams and, 
as such, they include both the positive- and negative-energy propagation of the intermediate 
particles. The nucleonic current is constructed consistently 
with the $NN$ potential in the BBS equation (3). 
The mesonic current consists of the $\eta\rho\rho$, $\eta\omega\omega$, 
and $\eta a_o\pi$ exchange contributions. The resonance current consists of the 
$S_{11}(1535)$, $P_{11}(1440)$ and $D_{13}(1520)$ resonances excited via the exchange of 
$\pi$, $\eta$, $\rho$ and $\omega$ mesons. Details of our model for the production current are
given in the appendix.

\section{Vector Meson Exchange Dominance Model}
In order to allow for a close comparison of the model 
described in the previous section 
with the models \cite{Wilkin,Gedalin,Santra} based on 
$\rho$ meson exchange dominance, we have also 
constructed a model in which the $S_{11}(1535)$
is excited through the exchange of vector mesons.
For this purpose, we follow Refs.\cite{Gedalin,Santra}
to define the vector meson couplings with $S_{11}(1535)$ by using
the following Lagrangian densites 
\begin{subequations}
\begin{eqnarray}
{\cal L}_{\omega NN^*}(x) & = & -
g_{\omega NN^*} \bar \psi_{N^*}(x) \gamma_5\gamma_\mu\omega^\mu(x) \psi_N(x) + h.c.  \ ,
\label{NR12vomega} \\
{\cal L}_{\rho NN^*}(x) & = & -
g_{\rho NN^*}\bar \psi_{N^*}(x) \gamma_5\gamma_\mu \vec \tau 
\cdot (\partial^\nu \vec \rho^\mu(x)) \psi_N(x) + h.c.  \ .
\label{NR12vrho}
\end{eqnarray}
\label{NR12v}
\end{subequations} 
Note that the above $\gamma_5 \gamma_\mu$ coupling,
which violates gauge invariance, is rather
different from the tensor coupling 
(see Eqs.(\ref{NR12omega},\ref{NR12rho})) used in our 
model described in the 
previous section and detailed in the appendix.

In addition to using the $\gamma_5 \gamma_\mu$ coupling in the $vNS_{11}(1535)$ vertex
($v=\rho,\omega$), here we 
assume the extreme case that $S_{11}(1535)$ is excited exclusively via the exchange 
of $\rho$ and $\omega$ mesons. Furthermore, for simplicity, we neglect all other resonance 
contributions in the resonance current. This is a reasonable simplification
since we find that the resonance current contributions 
apart from that due to $S_{11}(1535)$ are very small.
The nucleonic and mesonic currents are identical to the model described 
in the previous section. 
 We refer to this model as the vector meson 
exchange dominance model. 

 We find that this  model can 
also roughly reproduce the total cross section data by
choosing the coupling constants $g_{\rho NN^*} = -0.85$ 
and $g_{\omega NN^*} = -1.10$ in Eq.(\ref{NR12v}).
Overall, the values of these coupling constants, including the signs, are consistent 
with those used in Refs.\cite{Gedalin,Santra}, in spite of the fact that there the 
ISI and FSI are treated differently from the present work. 
We therefore can use this model to investigate the differences between
our full model described in section II and the previous 
works \cite{Wilkin,Gedalin,Santra} based on $\rho$ exchange dominance.

\section{Results}
In this section we shall present our results on the $\eta$ meson production in both the $pp$ and 
$pn$ collisions based on the models described in the previous sections. 
The parameters of the considered 
models are given explicitly in the appendix.
In short, the coupling constants and range of form
factors for meson-baryon-baryon vertices are chosen to be
consistent with the Bonn potential in conjunction with the values used in 
Refs.\cite{Nak2,Nak3} and the values extracted from Particle Data Group \cite{PDG}. 
Thus in our calculations there is not much freedom for adjusting parameters.

\begin{figure}
\vspace{15cm}
\includegraphics{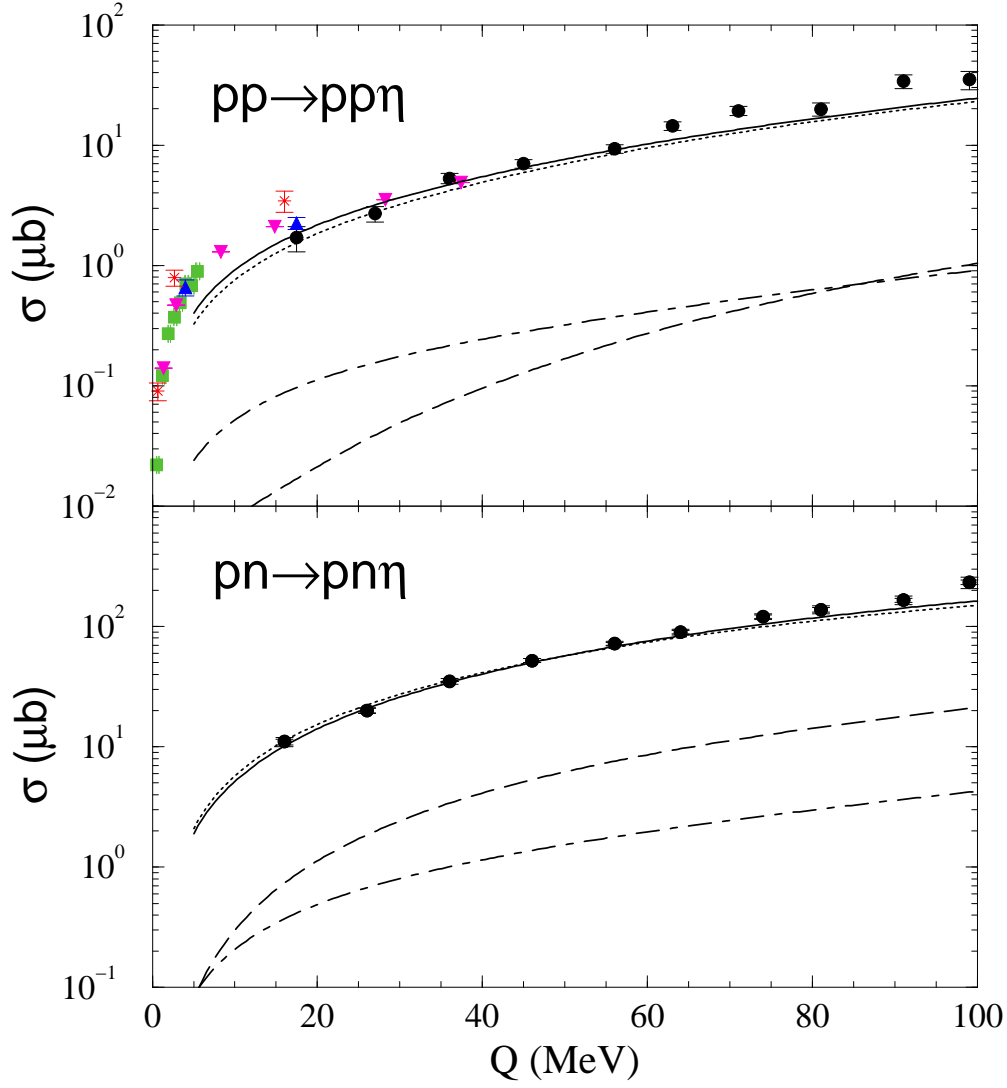}
\caption{Total cross sections for the $pp\rightarrow pp\eta$ (upper panel) and $pn\rightarrow pn\eta$ 
(lower panel) reactions as a function of excess energy within our model described in section II. 
 The dashed curves correspond to the nucleonic
current contribution while the dash-dotted curves to the mesonic current contribution; the dotted 
curves represent the resonance current contribution. The solid curves are the total contribution. 
The data are from Refs.\protect\cite{spes3eta1,spes3eta2,celsiuseta,cosy11eta,Calenp,Calenn}.}
\label{fig6}
\end{figure}

\begin{figure}[h]
\vspace{15cm}
\includegraphics{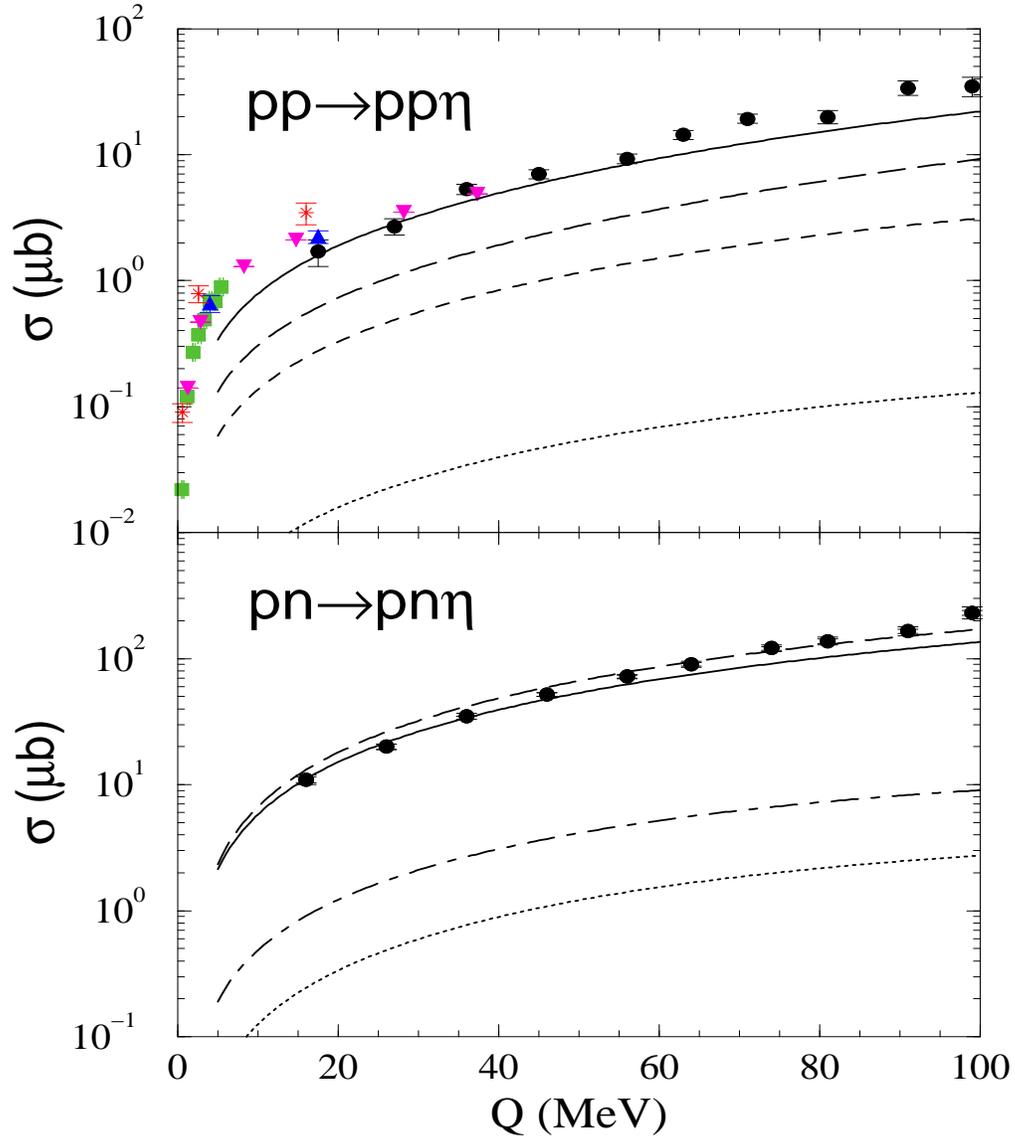}
\caption{Same as Fig.~\ref{fig6}, except that it shows the $S_{11}(1535)$ resonance contribution
only. The dashed curves correspond to the $\pi$ exchange contribution while the dash-dotted 
curves to the $\eta$ exchange contribution; the dotted curves represent the $\rho$ exchange 
contribution. The solid curves show the total contribution.} 
\label{fig7}
\end{figure}

The total cross sections as a function of excess energy predicted by 
our model (described in section II) are shown in Fig.~\ref{fig6}. 
The full results are the solid curves which are in general in
good agreement with  both the
data of $pp$(upper panel) and $pn$(lower panel) collisions. 
 For small excess 
energies, our $pp$ results underestimate the data. This is
usually attributed 
to the $\eta N$ FSI, which is not accounted for in the present model.
Note that the results for 
$pn \rightarrow pn\eta$ with excess energy $Q > 50\ MeV$, corresponding to an incident 
beam energy larger than $1.3\ GeV$, should be interpreted with caution, as no reliable $NN$ 
phase shift analyses for $T=0$ states exist at present for energies above $1.3\ GeV$ \cite{Igor}. 
To see the dynamical content of our model, we also show in Fig.3 the
results calculated from keeping only nuleonic current(dashed curves),
mesonic current(dash-dotted curves), and resonance current(dotted
curves).
We see that the total cross sections are obviously
 dominated by the resonance current, 
and more specifically by the strong $S_{11}(1535)$ resonance (see Fig.~\ref{fig7}). 
Our nucleonic current contributions (dashed curves) are much smaller than the resonance current 
contributions. This is rather different from
the findings of Ref.\cite{Pena}; there, instead of the 
resonance current,  the shorter range part of the nucleonic current 
gives a large contribution to the $pp\rightarrow pp\eta$ cross section.
It will be interesting to know whether their model can also give
a good description of $pn \rightarrow pn \eta$ data, as achieved 
here(lower panel of Fig.3).

To examine the differences between our model and previous work,
we show in Fig.~\ref{fig7} the results from calculations including only
the $S_{11}(1535)$ resonance(solid curves) contribution.
 Within our model, this resonance excitation
is due to the exchange of $\pi$, $\eta$, $\rho$, and $\omega$.
To see the relative importance between these different meson exchange
mechanisms, we also show in Fig~\ref{fig7} the results from
 $\pi$ exchange(dashed curves), $\eta$ exchange( dash-dotted curves), 
and $\rho$ exchange(dotted curves).
Although the $\omega$ exchange is included in the calculation, its 
contribution is not shown here
separately because it is much smaller than the $\rho$
 exchange contribution.   
As can be seen, the dominant contribution is due to $\pi$ exchange followed 
by $\eta$ exchange.
The $\rho$ exchange contribution is very small. Several observations are in order here:
\begin{itemize}  
\item[1)]
In contrast to the result of 
Refs.\cite{Gedalin,Santra}, our model, as given by Eq.(\ref{NR12rho}), 
does not allow the  $\gamma^5 \gamma^\mu $ coupling in the $\rho NN^*$ vertex 
for the considered negative parity $S_{11}(1535)$ resonance. Such a  
coupling would prevent us from determining the $\rho N N^*$
coupling from radiative decay $N^* \rightarrow \gamma N$ in the vector meson dominance 
model (VMD) as explained in the appendix, since
it violates the gauge invariance constraint which is an essential element of VMD.
The simplest way of satisfying gauge invariance 
is to omit the $\gamma^5\gamma^\mu$ coupling
and use only the tensor $\gamma^5\sigma^{\mu\nu}$ coupling, as given in
Eq.(\ref{NR12rho}). This choice of the $\rho NN^*$ coupling, combined with the 
corresponding coupling constant as given in Table \ref{tab1} -  
which is close to the low limit of the range determined from the measured 
radiative decay widths - leads to a very small 
$\rho$ exchange contribution to the cross section as shown in 
Fig.~\ref{fig7} (dotted curves). This is the main origin of the differences between our
results and that of  Refs.\cite{Gedalin,Santra} where
the $\rho NN^*$ vertex for $N^*=S_{11}(1535)$ 
is specified by the $\gamma^5\gamma^\mu$ coupling
(see also Fig.~\ref{fig9}). An 
alternative to avoid the gauge invariance problem while keeping the 
$\gamma^5\gamma^\mu$ term is to use
a vertex of the form 
$\gamma^5[\gamma^\mu q^2  - (m_{N^*}+ m_N)q^\mu]$ \cite{Riska}.
Pe\~na et al. \cite{Pena}, on the other hand, have used the vertex 
$\gamma^5[\gamma^\mu - (m_{N^*}+ m_N) q^\mu / m_\rho^2]$
in conjunction with the coupling constant determined from a quark model \cite{Riska}.
This vertex yields the same meson production amplitude as that of a pure 
$\gamma^5\gamma^\mu$ vertex while satisfying the gauge invariance constraint, although 
only in the on-shell limit ($q^2=m_\rho^2$).
 They found a significant contribution of $\rho$ exchange to the 
excitation of $S_{11}(1535)$ in $pp\rightarrow pp\eta$.
 Further experimental information is needed to
determine whether the $\gamma^5\gamma^\mu$ coupling is required
for vector meson exchange. 

\item[2)]
The $\eta$ exchange contribution is relatively large in the present calculation. In 
the case of $pp \rightarrow pp\eta$ its contribution to the cross section is about half of that 
due to the $\pi$ exchange. The $\eta$ exchange contribution is subject to a relatively large 
uncertainty which arises, apart from the introduction of the phenomenological form factors, from 
the uncertainty in the $\eta NN$ coupling strength as discussed in the appendix. The relatively 
large contribution of $\eta$ here results from using the $\eta NN$ coupling constant of 
$g_{\eta NN} = 6.14$, as used in the construction of the Bonn $NN$ interaction \cite{MHE87}. 
This is close to the upper limit of the range determined empirically as mentioned in the appendix.
However, the $\eta$ meson exchange in the Bonn potential \cite{MHE87} represents the exchange of 
a $(J^P,T)=(0^-,0)$ quantum number and not necessarily of a genuine $\eta$ meson. 
On the other hand, the value 
of $g_{\eta NN} = 6.14$ together with the $\eta -\eta^\prime$ mixing angle of 
$\theta _{P}\simeq -9.7{{}^\circ}$, as suggested by the quadratic mass formula, and the $\pi NN$ 
coupling constant of $g_{\pi NN }=13.45$ leads through SU(3) flavor symmetry to the ratio 
$D/F\simeq 1.43$. This is not too far from the value of $D/F\cong 1.73$ extracted from a systematic 
analysis of semileptonic hyperon decays \cite{DFratio}. Anyway, in the present 
calculation for $pp \rightarrow pp\eta$, the 
$\eta$ exchange interferes constructively with the dominant $\pi$ exchange contribution, yielding the 
total contribution as shown by the solid line in Fig.~\ref{fig7}. For $pn \rightarrow pn\eta$, the 
$\eta$ exchange interferes constructively with the $\pi$ exchange in the $T=1$ channel (as in the case 
of $pp \rightarrow pp\eta$), but destructively in the $T=0$ channel due to the isospin factor $-3$ in 
the $\pi$ exchange amplitude. 
\item[3)]
The correct description of both  
$pp \rightarrow pp\eta$ and $pn \rightarrow pn\eta$ reactions 
depends not only on the isospin 
factors associated with the isovector and isoscalar meson exchange but, also on a 
delicate interplay between
the $NN$ FSI and ISI in each partial wave. 
While the $NN$ FSI enhances the total cross 
section, the $NN$ ISI has an opposite effect (see discussion in section II). In this connection, 
we mention that in Ref.\cite{Wilkin} the reduction factor due to the $NN$ ISI is estimated to be 
about $\lambda_{(^3P_0)}=(0.77)^2=0.59$ due to the $^3P_0$ state and $\lambda_{(^1P_1)}=(0.73)^2=0.53$ 
due to $^1P_1$. In our calculation, however, the corresponding reduction factors
are about $\lambda_{(^3P_0)}=0.19$ and $\lambda_{(^1P_1)}=0.27$ near the threshold energy. This large 
discrepancy between the results of 
Ref.\cite{Wilkin} and ours is due to the fact that, whereas our reduction factor is given by 
Eq.(\ref{isieff}), the reduction factor in Ref.\cite{Wilkin} is given by $\lambda_i \equiv \eta_i^2 = 
(e^{-Im(\delta_i)})^2$. We argue that the latter formula is inappropriate for estimating the effect of 
the $NN$ ISI since it exhibits a pathological
 feature: namely, when the absorption is 
maximum ($\eta_i=0$), this formula yields $\lambda_i = 0$, implying the total absence of the $NN$ 
elastic channel and thus not allowing the production reaction to occur. However, scattering theory 
tells us that when the absorption cross section is maximum, the corresponding elastic cross 
section does not vanish, but is 1/4 of the absorption cross 
section. Note that this feature is present in Eq.(\ref{isieff}). Furthermore, the authors of 
Ref.\cite{Wilkin} apparently have identified incorrectly the inelasticity $\eta_i$ with 
$\cos^2(\rho_i)$, where $\rho_i$ is one of the two parameters (the other is the phase shift) given in 
Ref.\cite{SAID}. The phase shift parametrization given in Ref.\cite{SAID} differs from the standard 
Stapp parametrization. It is obvious that with a more appropriate 
estimate of the reduction factor $\lambda_i$ as given by Eq.(\ref{isieff}) the result of 
Ref.\cite{Wilkin} would underpredict considerably the cross section data.
\end{itemize}

\begin{figure}[h]
\vspace{15cm}
\includegraphics{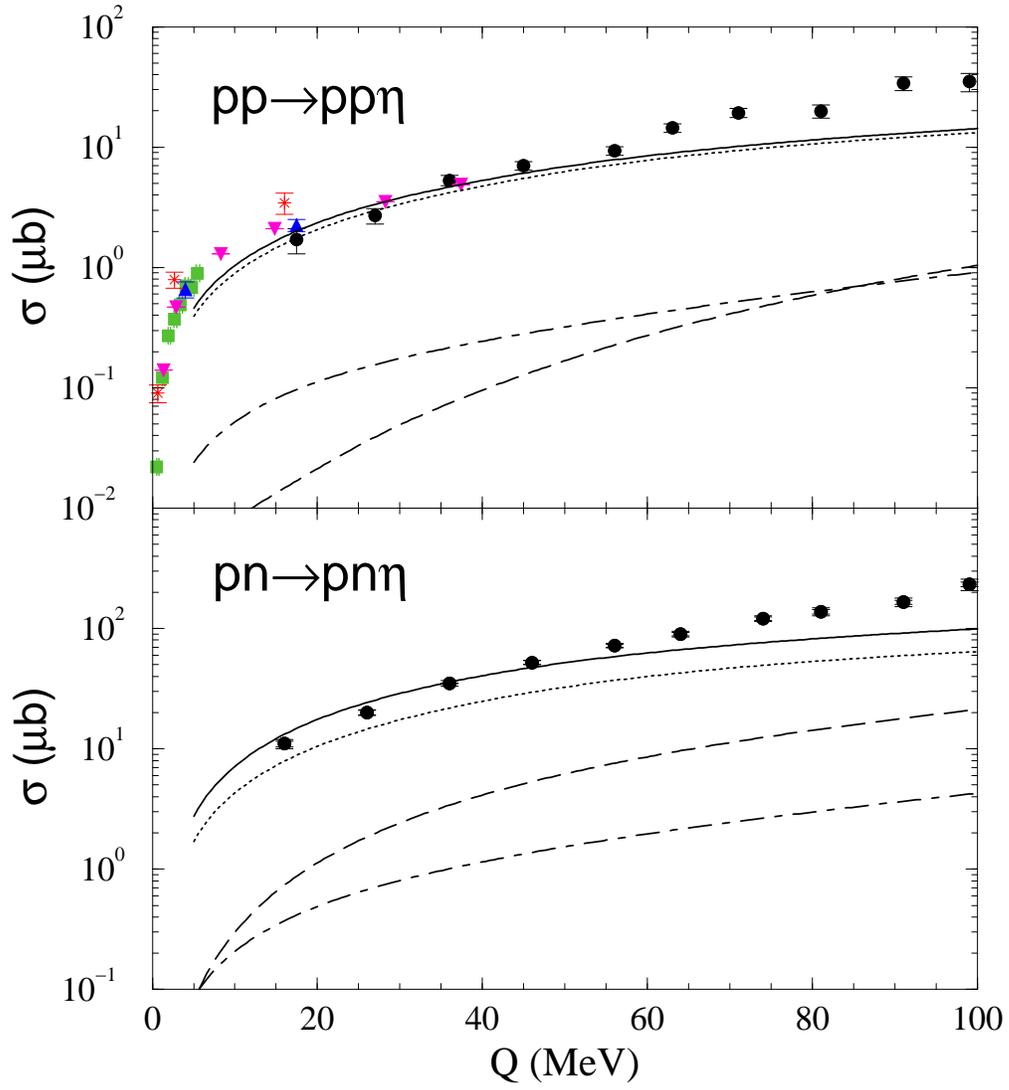}
\caption{Same as Fig.~\ref{fig6} but using the vector meson exchange dominance described in section III.}
\label{fig8}
\end{figure}

\begin{figure}[h]
\vspace{15cm}
\includegraphics{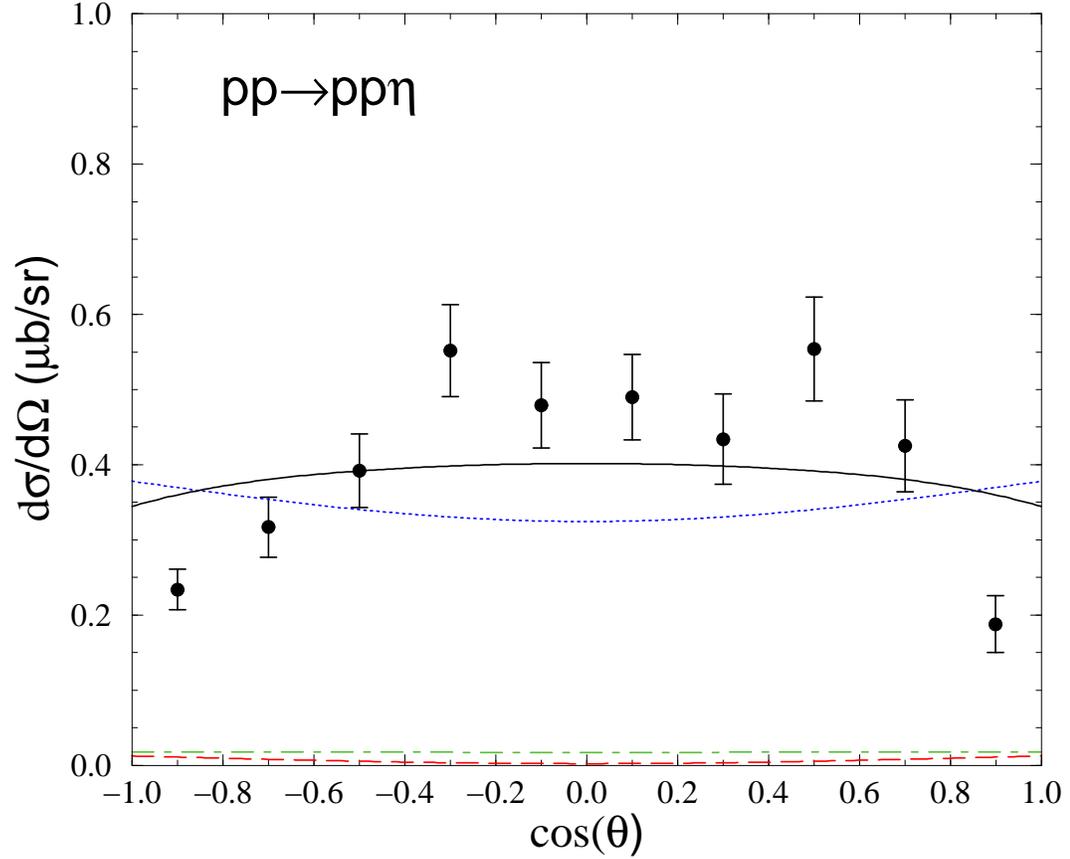}
\caption{Angular distribution of the emitted $\eta$ meson in the c.m. frame of the total
system at an excess energy of $Q=37\ MeV$. The dashed curve corresponds to
the nucleonic current contribution while the dash-dotted curve to the mesonic current 
contribution; the dotted curves represent the resonance current contribution. The solid curve 
show the total contribution. The data are from Ref.\protect\cite{Calenh}.}
\label{fig9}
\end{figure}

We now turn to exploring the vector meson exchange dominance model
described in section III. This model does not have the $\pi$ and $\eta$
exchange mechanisms for exciting the $S_{11}(1353)$ resonance. The
nucleonic and mesonic currents of our full model are kept.
With the coupling constants $g_{\rho NN^*} = -0.85$ and $g_{\omega NN^*} = -1.10$
in Eq.(\ref{NR12v}), we can describe 
both the $pp\rightarrow pp \eta$ and $pn\rightarrow pn\eta$ data. 
The results are the solid curves in
Fig.~\ref{fig8}. In the same figures we also show the contributions
from the nucleon resonance (dotted curves) and the nucleonic (dashed curves)
and mesonic (dash-dotted curves) currents.
Although the total cross section is
underpredicted for excess energies $Q > 60 \ MeV$, it is 
interesting that these 
results are in line with the findings of 
Refs.\cite{Wilkin,Gedalin,Santra}. 

\begin{figure}[h]
\vspace{15cm}
\includegraphics{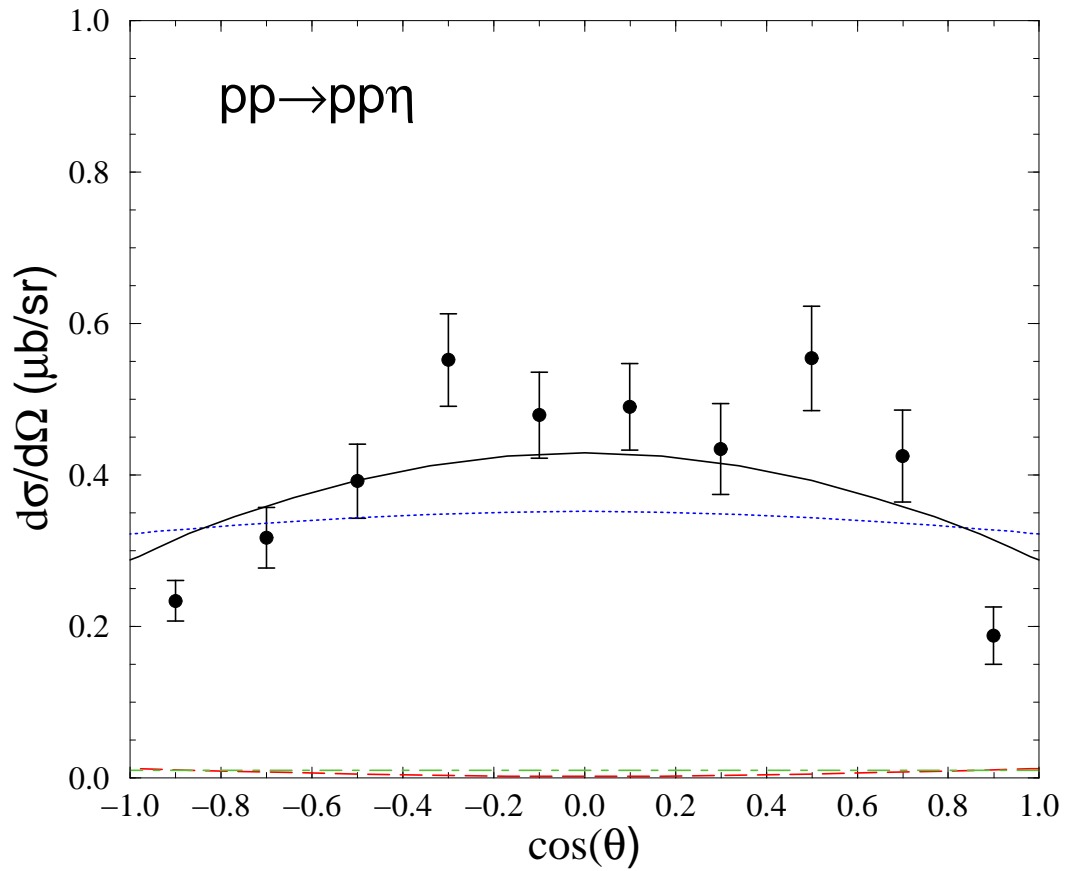}
\caption{Same as Fig.~\ref{fig9} but using the vector meson exchange dominance of section III.}
\label{fig10}
\end{figure}

The results presented above indicate that the total cross section
data alone cannot distinguish two different meson exchange mechanisms
for the excitation of $S_{11}(1535)$ resonance. It is therefore necessary
to consider more exclusive observables. 
Fig.~\ref{fig9} shows the angular distribution(solid curves) of
 $pp \rightarrow pp\eta$ predicted by our model
(described in section II) at an excess energy of $Q=37\ MeV$.
The data from Ref.\cite{Calenh} are also shown. Again, the resonance 
contribution (dotted curve) dominates the cross section. As pointed out in Ref.\cite{Wilkin}, the 
shape of the angular distribution of the latter contribution bends upwards at the forward and 
backward angles due to the $\pi$ exchange dominance in the $S_{11}(1535)$ resonance contribution.
However, due to interference with the nucleonic (dashed) and mesonic (dash-dotted) currents, the 
shape of the resulting angular distribution (solid curve) is inverted with respect to that of the 
resonance current contribution alone. As one can see, although the overall magnitude is rather well 
reproduced, the rather strong angular dependence exhibited by the data
 is not reproduced by the model described in section II.
At this point one might argue that the excitation mechanism 
of the $S_{11}(1535)$ resonance as given 
by our model is not correct and that, indeed, the $\rho$ meson 
exchange is the 
dominant contribution, as has been claimed in Ref.\cite{Wilkin}.
This can be studied by considering the predictions of
our vector meson exchange dominance model described in section III.
The angular distribution predicted by this model is shown in 
 Fig.~\ref{fig10}. 
Here we see that the shape of the calculated 
angular distribution(solid curve) is in better agreement with
the data, although the strong angular 
dependence exhibited by the data - which shows contributions of higher partial waves than $L=1$ - is
not quite reproduced. Judging from the level of 
agreement between the two predictions
and the data, one cannot discard our model in which the
$S_{11}(1535)$ is mainly excited by $\pi$ and $\eta$ exchange
 in favor of the 
$\rho$ exchange dominance model. In this connection, we 
mention that new data from COSY which will become available soon shows a flat angular 
distribution \cite{Kilian}.

\begin{figure}[h]
\vspace{15cm}
\includegraphics{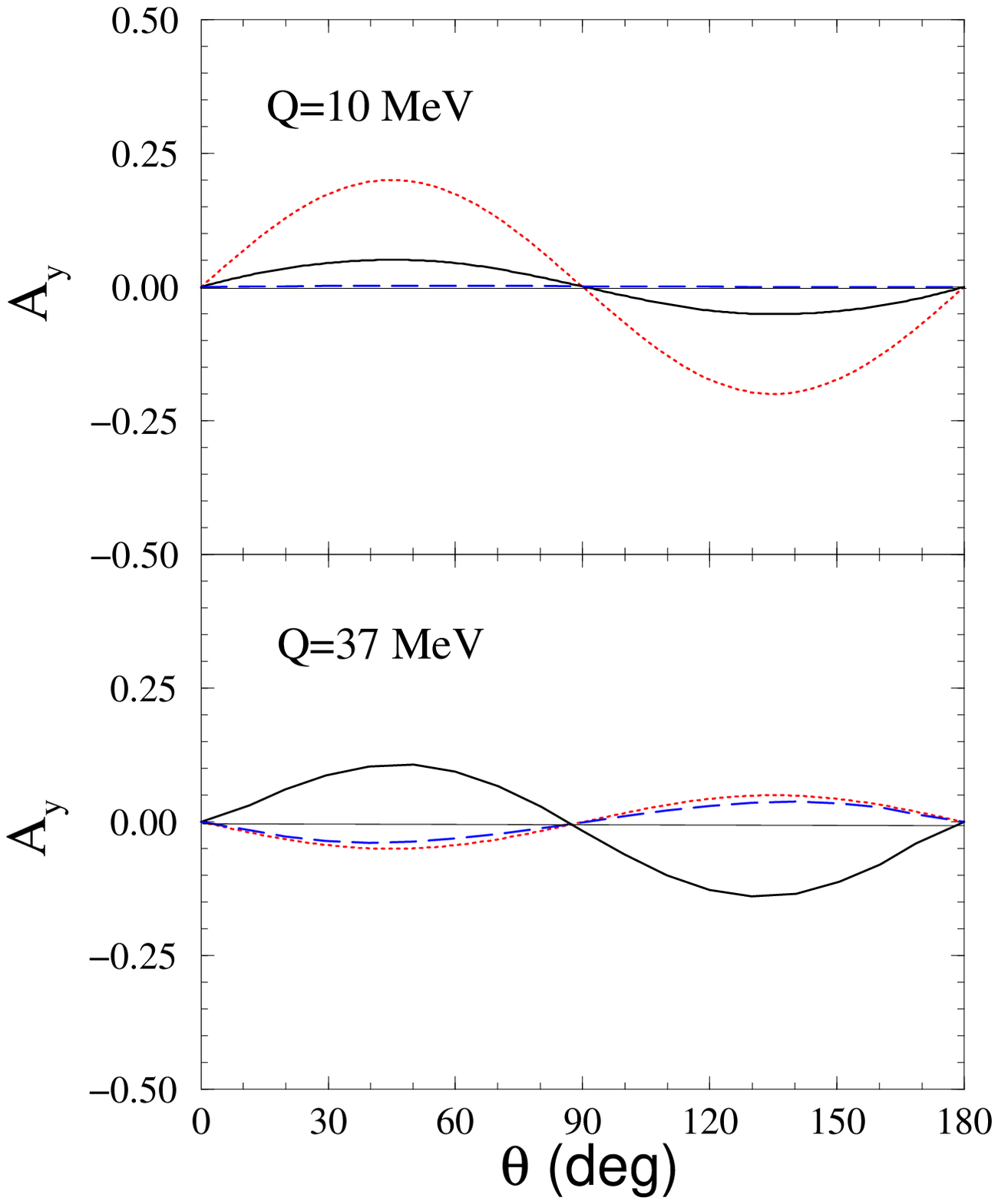}
\caption{Analyzing power for the reaction $pp\rightarrow pp\eta$ as a function of emission-angle
of $\eta$ in the c.m. frame of the total system at an excess energy of $Q=10\ MeV$ (upper
panel) and $Q=37\ MeV$ (lower panel). The dotted curve corresponds to the case of $\rho$ exchange 
dominance according to Ref.\protect\cite{Wilkin}. The solid curve corresponds to the 
full model calculation described in section II. The dashed curve is the prediction assuming vector 
meson exchange dominance as described in section III.}
\label{fig11}
\end{figure}

From the above considerations, we conclude that, at present, the excitation mechanism of the 
$S_{11}(1535)$ resonance in $NN$ collisions is still an open question. Indeed, we have just offered 
a scenario which is as good as the $\rho$ exchange dominance model in
reproducing the available data. 
It is therefore of special interest to seek a way to disentangle these possible scenarios. In this
connection, spin observables may potentially help resolve this issue. 
As an example, we present in Fig.~\ref{fig11} the analyzing
power at $Q=10\ MeV$ 
(upper panel) and $Q=37\ MeV$ (lower panel). The predictions of the
present model are shown as the solid curves, whereas the predictions assuming 
vector meson exchange dominance for exciting the $S_{11}(1535)$ resonance are shown as the dashed curves. 
The different features exhibited by the two scenarios for the excitation mechanism of 
the $S_{11}(1535)$ is evident. 
According to Ref.\cite{Wilkin}, the $\rho$ exchange contribution is expected to lead to an analyzing 
power given by
\begin{equation}
A_y = A_y^{max}\sin(2\theta) \ , 
\label{rhoAy}
\end{equation}
where $A_y^{max}$ is positive for low excess energies, peaking at $Q \approx 10\ MeV$ and becoming 
negative for excess energies $Q > 35\ MeV$. The corresponding results are also shown in 
Fig.~\ref{fig11} as the dotted curves. Although Eq.(\ref{rhoAy}) gives rise to a larger analyzing power 
at $Q = 10 \ MeV$, it is interesting to see that, at $Q=37 \ MeV$, it yields a result that nearly 
coincides with the prediction of the vector meson exchange dominance given in section III.

\section{Summary}
The production of $\eta$ mesons in $NN$ collisions has been discussed within a relativistic meson 
exchange model of hadronic interactions, where the production current has been constructed consistently 
with the $NN$ FSI used. Special emphasis has been paid to investigate the possible excitation 
mechanisms of the $S_{11}(1535)$ resonance, which is currently a subject of debate. It has been shown 
that, not only the vector meson dominance for exciting the $S_{11}(1535)$ as advocated by Wilkin and
collaborators \cite{Wilkin}, but also the excitation mechanism of this resonance mainly via exchange of 
$\pi$ and $\eta$ mesons can describe the existing data in both the $pp$ and $pn$ collisions equally well. 
We have found that the analyzing power may offer an opportunity to disentangle these reaction mechanisms.

A consistent description of the meson production reaction in $pp$ and $pn$ collisions is not a trivial
task. As we have seen, this depends not only on the different isospin factors in the production current 
which change the relative importance of different reaction mechanisms from $pp$ to $pn$ collisions, but 
also on a delicate interplay between the $NN$ FSI and ISI. It is clear that the $pp\rightarrow ppM$ 
and $pn\rightarrow pnM$ as well as the $pn\rightarrow dM$ reaction should be investigated in a consistent
way. Also more exclusive observables than the total cross section such as the spin observables should be 
studied.

Finally, we emphasize that the results
 presented in this paper
should be interpreted with 
caution. The reason for this is that, as mentioned before, 
the $NN$ ISI is only accounted for 
using the on-shell approximation. While this may be a reasonable
 approximation for calculating 
cross sections, it may introduce rather large uncertainties in the
 calculated spin observables. Efforts to improve this will be
published elsewhere. 

\vskip 0.5cm
{\bf Acknowledgment: }
We would like to thank V. Baru, J. Durso, J. Haidenbauer 
and C. Hanhart for valuable discussions. We also thank W. G. Love for a careful reading
of the manuscript.
This work is supported by Forschungszentrum-J\"ulich, Contract 
No 41445282(COSY-58) and by 
U.S. Department of Energy, Nuclear Physics Division, Contract
No. W-31-109-ENG-38.

\section{Appendix: Production Currents}

The $\eta$ meson production current consists of nucleonic, mesonic and resonance 
currents as shown in Eq.(\ref{curr}) and illustrated diagrammatically in Fig.~~\ref{fig1}. 
In the following subsections we construct each of these currents. A general remark to be
mentioned here which applies to all of the currents constructed in the following subsections
is that, as a consequence of using a three dimensional reduction of the Bethe-Salpeter 
equation in evaluating the total amplitude in Eq.(\ref{ampl0}), the 
time components of 
the intermediate particles involved in the production current suffer from an ambiguity in 
their definitions. In order to be consistent with the $NN$ interaction used in the present 
work, which has been constructed by using the BBS three dimensional reduction, the time
component of the four-momentum of a virtual meson at the
$MNN$ vertex is taken to be $q_o = \varepsilon(l) - 
\varepsilon(l')$, with $\varepsilon(l)$ and $\varepsilon(l')$ 
denoting respectively the energies of the
nucleon before and after the emission of the virtual meson.
$\varepsilon(l)\equiv \sqrt{l^2 + m_N^2}$. The time component of the intermediate baryon 
in the nucleonic and resonance currents are taken to be $p_o = \omega(k) + \varepsilon(p')$ 
at the $B \rightarrow M + N$ vertex, while at the $N \rightarrow M + B$ vertex we take $p_o 
= \varepsilon(p') - \omega(k)$. Here, $\omega(k)$ stands for the energy of the meson produced 
in the final state.

\subsection{The Nucleonic Current}

The nucleonic current is defined as
\begin{equation}
J_{nuc} = \sum_{j=1,2}\left ( \Gamma_j iS_j U + U iS_j \Gamma_j \right ) \ ,
\label{nuc_cur}
\end{equation}
with $\Gamma _{j}$ denoting the $\eta NN$ vertex and $S_{j}$ the nucleon 
(Feynman) propagator for nucleon $j$. The summation runs over the two interacting 
nucleons, 1 and 2. $U$ stands for the meson-exchange $NN$ potential. It is, in 
principle, identical to the potential $V$ appearing in the $NN$ scattering equation, 
except that here meson retardation effects (which are neglected in the potential 
entering in Eq.(\ref{scateqn})) are kept as given by the Feynman prescription.

The structure of the $\eta NN$ vertex, $\Gamma _{j}$, in
Eq.(\ref{nuc_cur}) is derived from the Lagrangian density
\begin{equation}
{\cal L}(x) =  - g_{\eta NN} \bar\psi_N(x) \gamma_5 
\left( i\lambda + \frac {1-\lambda} {2m_N}\gamma^\mu \right) \psi_N(x)
\partial_\mu \eta(x) \ ,
\label{NNps}
\end{equation}
where $g_{\eta NN}$ denotes the $\eta NN$ coupling constant and 
$\lambda $ is the parameter controlling the pseudoscalar(ps) - pseudovector(pv) 
admixture. $\eta(x)$ and $\psi_N(x)$ stand for the $\eta$ and 
nucleon field, respectively, and $m_{N} $ denotes the nucleon mass.

The coupling constant $g_{\eta NN}$ is poorly known at present. The empirical
values for $g_{\eta NN}$ range anywhere from 1 to 7 
\cite{Bernard,MHE87,Benmer,Tiator}. The values extracted from the $\eta$ 
photoproduction analysis tend to be in the low side of 
this range \cite{Tiator}, while a value of
$g_{\eta NN} = 6.14$ has been used
in the NN scattering analysis by the Bonn group \cite{MHE87}.
 In the present work we use the value of 
$g_{\eta NN}=6.14$, consistent with the $NN$ potential $V$ appearing in 
Eq.(\ref{scateqn}). Also, we take the pure pseudovector coupling, $\lambda = 0$. 

The $\eta NN$ vertex derived from Eq.(\ref{NNps}) should be provided with an 
off-shell form factor. Following Ref.\cite{Nak3}, we
 associate each nucleon leg with a form factor of the
following form 
\begin{equation}
F_N(p^2) = \frac{\Lambda_N^4} {\Lambda_N^4 + (p^2-m_N^2)^2}  \ ,
\label{formfactorN}
\end{equation}
with $\Lambda_N=1.2 \ GeV$. $p^{2}$ denotes the four-momentum squared of either the 
incoming or outgoing 
off-shell nucleon. We also introduce the form factor given by Eq.(\ref{formfactorN}) 
at those $MNN$ vertices appearing next to the $\eta$-production vertex, 
where the (intermediate) nucleon and the exchanged mesons are off their mass shell 
(see Fig.~\ref{fig1}). Therefore, the corresponding form 
factors are given by the product $F_{N}(p^{2})F_{M}(q^{2})$, where $M$ stands for 
each of the exchanged mesons between the two interacting nucleons. The form factor 
$F_{M}(q^{2})$ accounts for the exchanged meson being off-shell and is 
taken to be consistent  with the 
considered Bonn $NN$ potential used for generating
the final NN scattering wavefunction.

\subsection{The Resonance Current}

The production of $\eta$ mesons in $NN$ collisions is thought to occur predominantly through 
the excitation (and de-excitation) of the $S_{11}(1535)$ resonance, to which the $\eta$ meson 
couples strongly. In the present work, we also consider 
the $P_{11}(1440)$ and $D_{13}(1520)$ resonances. 

The resonance current is composed of the spin-1/2 and spin-3/2 resonance contributions 
\begin{equation}
J_{res} = J^{(1/2)}_{res} + J^{(3/2)}_{res} \ .
\label{res_cur}
\end{equation}
The spin-1/2 resonance current, in analogy to the nucleonic current, is written as
\begin{equation}
J^{(1/2)}_{res}  =  \sum_{j=1,2}\sum_{N^*}\left ( \Gamma_{\eta jN^*} iS_{N^*} U_{N^*} + 
\tilde U_{N^*} iS_{N^*} \Gamma_{\eta jN^*} \right ) \ .
\label{res12_cur}
\end{equation}
Here $\Gamma _{\eta jN^*}$ stands for the $\eta NN^*$ vertex 
function involving the nucleon $j$. $S_{N^*}(p) = ({\mbox{$ \not \! p$}} + m_{N^*}) / 
(p^2 - m_{N^*}^2 + im_{N^*}\Gamma_{N^*})$ is the $N^*$ resonance propagator, with 
$m_{N^*}$ and $\Gamma_{N^*}$ denoting the mass and width of the resonance, 
respectively. The summation runs over the two interacting nucleons, $j=1$ and $2$, 
and also over the spin-1/2 resonances considered, i.e., $N^* = S_{11}(1535)$ and 
$P_{11}(1440)$. In the above equation $U_{N^*}$ ($\tilde U_{N^*}$) stands for the 
$NN\rightarrow NN^*$ ($NN^*\rightarrow NN$) meson-exchange transition potential. 
It is given by
\begin{equation}
U_{N^*} = \sum_{M=\pi, \eta} \Gamma_{MNN^*}(q) i\Delta_M(q^2)  \Gamma_{MNN}(q)
       + \sum_{M=\rho, \omega} \Gamma^\mu_{MNN^*}(q) iD_{\mu\nu (M)}(q)
\Gamma^\nu_{MNN}(q)   \ ,
\label{transpot12}
\end{equation}
where $\Delta_M(q^2)$ and $D_{\mu\nu (M)}(q)$ denote the (Feynman) propagator of the 
exchanged pseudoscalar and vector meson, respectively.
$\Gamma_{MNN}(q)$ and $\Gamma^\mu_{MNN}(q)$ denote the pseudoscalar and vector $MNN$ 
vertex, respectively. These vertices are taken consistently with the $NN$ potential 
$V$ appearing in Eq.(\ref{scateqn}) except for the type of coupling at the $\pi NN$ 
vertex and the $\omega NN$ coupling constant. Following the discussion in 
Ref.\cite{Nak3}, we use the pv-coupling ($\lambda = 0$) instead of the ps-coupling 
($\lambda =1$) at the $\pi NN$ vertex. Also, following the discussions in 
Refs.\cite{Nak1,Nak2,Nak3}, the value of the $\omega NN$ coupling constant is taken to 
be $g_{\omega NN} = 11.76$. These exceptions apply to the currents $J_{res}^{(3/2)}$ and 
$J_{mec}$ as well. Analogous expression to Eq.(\ref{transpot12}) holds for $\tilde U_{N^*}$.

Following Ref.\cite{Benmer}, the transition vertices $\Gamma_{MNN^*}$ and 
$\Gamma^\mu_{MNN^*}$ in Eqs.(\ref{res12_cur},\ref{transpot12}) for spin-1/2 
resonances are obtained from the interaction Lagrangian densities 
\begin{subequations}
\begin{eqnarray}
{\cal L}^{(\pm)}_{\eta NN^*}(x) & = & \mp g_{\eta NN^*}
\bar \psi_{N^*}(x) \left\{ \left[ i\lambda\Gamma^{(\pm)} + 
\left(\frac {1 - \lambda} {m_{N^*}\pm m_N}\right)\Gamma^{(\pm)}_\mu 
\partial^\mu \right] \eta(x) \right\} \psi_N(x) + h.c. \ , 
\label{NR12eta} \\
{\cal L}^{(\pm)}_{\pi NN^*}(x) & = & \mp g_{\pi NN^*}
\bar \psi_{N^*}(x) \left\{ \left[ i\lambda\Gamma^{(\pm)} + 
\left(\frac {1 - \lambda} {m_{N^*}\pm m_N}\right)\Gamma^{(\pm)}_\mu \partial^\mu
\right] \vec\tau \cdot \vec\pi(x) \right\} \psi_N(x) + h.c. \  ,
\label{NR12pi} \\
{\cal L}^{(\pm)}_{\omega NN^*}(x) & = & 
\left(\frac {g_{\omega NN^*}} {m_{N^*}+m_N}\right)
\bar \psi_{N^*}(x) \Gamma^{(\mp)} \sigma_{\mu\nu}
(\partial^\nu \omega^\mu(x)) \psi_N(x) + h.c.  \ ,
\label{NR12omega} \\
{\cal L}^{(\pm)}_{\rho NN^*}(x) & = & 
\left(\frac {g_{\rho NN^*}} {m_{N^*}+m_N}\right)
\bar \psi_{N^*}(x) \Gamma^{(\mp)} \sigma_{\mu\nu} \vec \tau
\cdot (\partial^\nu \vec \rho^\mu(x)) \psi_N(x) + h.c.  \ ,
\label{NR12rho}
\end{eqnarray}
\label{NR12M}
\end{subequations}
where $\vec\pi(x)$, $\omega^\mu(x)$, $\vec\rho^\mu(x)$ and $\psi_{N^*}(x)$ denote the 
$\pi$, $\omega$, $\rho$ and spin-1/2 nucleon resonance fields, respectively. The upper 
and lower signs refer to the even(+) and odd(-) parity resonance, respectively. The operators 
$\Gamma^{(\pm)}$ and $\Gamma^{(\pm)}_\mu$ in the above equations are given by
\begin{eqnarray}
\Gamma^{(+)} = \gamma_5 & \ , & \Gamma^{(+)}_\mu = \gamma_5 \gamma_\mu \nonumber \\
\Gamma^{(-)} = 1        & \ , & \Gamma^{(-)}_\mu = \gamma_\mu \ .
\label{NRpsa}
\end{eqnarray}
The parameter $\lambda$ in Eqs.(\ref{NR12eta},\ref{NR12pi}) controls the admixture of 
the two types of couplings: ps ($\lambda=1$) and pv ($\lambda=0$) in the case of an even 
parity resonance and, scalar ($\lambda=1$) and vector ($\lambda=0$) in the case of 
an odd parity resonance. On-shell, both choices of the parameter $\lambda$ are 
equivalent. In this work we take $\lambda=0$. Note that we have not allowed the coupling 
$\Gamma^{(\mp)}_\mu$ in Eqs.(\ref{NR12omega},\ref{NR12rho}) in contrast to 
Refs.\cite{Gedalin,Santra}. Unlike the $vNN$ vertex ($v=$vector meson), this coupling 
at the $vNN^*$ vertex prevents us from estimating its 
strength using the VMD because of the violation of
gauge invariance. Although gauge invariant 
vertices which include the $\Gamma^{(\mp)}_\mu$ coupling can be constructed \cite{Riska}, 
we have omitted this coupling in the present work for simplicity.

Similar to the case of spin-1/2 resonances, the spin-3/2 resonance current is written as
\begin{equation}
J^{(3/2)}_{res}  =  \sum_{j=1,2}\sum_{N^*}\left ( \Gamma^\mu_{\eta jN^*} iS_{\mu\nu (N^*)} 
U^\nu_{N^*} + 
\tilde U^\mu_{N^*} iS_{\mu\nu (N^*)} \Gamma^\nu_{\eta jN^*} \right ) \ .
\label{res32_cur}
\end{equation}
Here $\Gamma^\nu_{\eta jN^*}$ stands for the $\eta NN^*$ vertex function involving 
the nucleon $j$. $S_{\mu\nu (N^*)}(p) = ({\mbox{$ \not \! p$}} + m_{N^*})
\left\{- g_{\mu\nu} + \gamma_\mu\gamma_\nu/3
+ (\gamma_\mu p_\nu - p_\mu\gamma_\nu)/3m_{N^*} + 2p_\mu p_\nu / 3m^2_{N^*} \right\}
/ (p^2 - m_{N^*}^2 + im_{N^*}\Gamma_{N^*})$ 
is the spin-3/2 Rarita-Schwinger propagator. The summation runs over the two interacting 
nucleons, $j=1$ and $2$, and also over the spin-3/2 resonances considered, i.e., 
$N^* = D_{13}(1520)$ in the present work. In the above equation $U^\mu_{N^*}$ 
($\tilde U^\mu_{N^*}$) stands for the $NN\rightarrow NN^*$ ($NN^*\rightarrow NN$) 
meson-exchange transition potential. It is given by
\begin{equation}
U^\mu_{N^*} = \sum_{M=\pi, \eta} \Gamma^\mu_{MNN^*}(q) i\Delta_M(q^2)  \Gamma_{MNN}(q)
       + \sum_{M=\rho, \omega} \Gamma^{\mu\lambda}_{MNN^*}(q) iD_{\lambda\nu (M)}(q)
\Gamma^\nu_{MNN}(q)   \ ,
\label{transpot32}
\end{equation}
where $\Gamma^\mu_{MNN^*}(q)$ and $\Gamma^{\mu\lambda}_{MNN^*}(q)$ denote the 
pseudoscalar and vector $MNN^*$ vertex, respectively. An analogous expression to 
Eq.(\ref{transpot32}) holds for $\tilde U^\mu_{N^*}$.

The $MNN^*$ vertices involving spin-3/2 nucleon resonances in 
Eqs.(\ref{res32_cur},\ref{transpot32}) are obtained from the Lagrangian densities 
\cite{Benmer}
\begin{subequations}
\begin{eqnarray}
{\cal L}^{(\pm)}_{\eta NN^*}(x) & = & \left(\frac {g_{\eta NN^*}} {m_\eta}\right)
\bar \psi_{N^*}^\mu(x) \Theta_{\mu\nu}(z) \Gamma^{(\mp)} \psi_N(x)  
\partial^\nu \eta(x) + h.c. \ ,
\label{NR32eta} \\
{\cal L}^{(\pm)}_{\pi NN^*}(x) & = & \left(\frac {g_{\pi NN^*}} {m_\pi}\right)
\bar \psi_{N^*}^\mu(x) \Theta_{\mu\nu}(z) \Gamma^{(\mp)} \vec\tau \psi_N(x)
\cdot \partial^\nu \vec\pi(x) + h.c. \ ,
\label{NR32pi} \\
{\cal L}^{(\pm)}_{\omega NN^*}(x) & = & 
\mp i \left(\frac {g_{\omega NN^*}^{(1)}} {2m_N}\right)
\bar \psi_{N^*}^\mu(x) \Theta_{\mu\nu}(z) \Gamma^{(\pm)}_\lambda 
\psi_N(x) \omega^{\lambda\nu}(x) \nonumber \\ 
& - &  \left(\frac {g_{\omega NN^*}^{(2)}} {4m_N^2}\right) \left( 
\partial_\lambda \bar \psi_{N^*}^\mu(x) 
\Theta_{\mu\nu}(z) \Gamma^{(\pm)} \psi_N(x) \right)
\omega^{\lambda\nu}(x)  + h.c. \  ,
\label{NR32omega} \\
{\cal L}^{(\pm)}_{\rho NN^*}(x) & = & 
\mp i \left(\frac {g_{\rho NN^*}^{(1)}} {2m_N}\right) \bar \psi_{N^*}^\mu(x) 
\Theta_{\mu\nu}(z) \Gamma^{(\pm)}_\lambda \vec\tau \psi_N(x) \cdot 
\vec\rho^{\lambda\nu}(x) \nonumber \\
& - &  \left(\frac {g_{\rho NN^*}^{(2)}} {4m_N^2}\right) \left( 
\partial_\lambda \bar \psi_{N^*}^\mu(x) 
\Theta_{\mu\nu}(z) \Gamma^{(\pm)} \vec\tau \psi_N(x) \right) \cdot
\vec\rho^{\lambda\nu}(x) + h.c. \  ,
\label{NR32rho} 
\end{eqnarray}
\label{NR32M}
\end{subequations}
where $\Theta_{\mu\nu}(z) \equiv g_{\mu\nu} - (z+1/2)\gamma_\mu \gamma_\nu$. In order to
reduce the number of parameters, we take $z=-1/2$ in the present work.
$\omega^{\lambda\nu}(x)\equiv \partial^\lambda\omega^\nu(x) - \partial^\nu\omega^\lambda(x)$ and 
$\vec\rho^{\lambda\nu}(x)\equiv \partial^\lambda\vec\rho^\nu(x) - \partial^\nu\vec\rho^\lambda(x)$.

The coupling constants $g_{MNN^*}$ used in the present work are displayed in Table.\ref{tab1}.
They are determined from the centroid values of the extracted decay widths (and masses) of the 
resonances from Ref.\cite{PDG} whenever available. Those involving vector mesons, are estimated 
from the corresponding radiative decay width in conjunction with the VMD.
In order to reduce the 
number of free parameters, the ratio of the $vNN^*$ ($v=\rho,\omega$) coupling constants for the 
spin-3/2 $D_{13}(1520)$ resonance has been fixed to be $g_{vNN^*}^{(1)}/g_{vNN^*}^{(2)}=-3$.  
This is not too far from the ratio of $g_{\gamma NN^*}^{(1)}/g_{\gamma NN^*}^{(2)}=-2.1$ 
for $N^*=P_{33}(1232)$ extracted from the ratio of $E2/M1 \cong -2.5\%$ as determined from pion 
photoproduction measurements \cite{MAINZ}.
As for the coupling constant $g_{vNN*}$ ($v = \rho , \omega$) for $N^*=S_{11}(1535)$, we use 
a value close to the lower limit of the range determined from the radiative decay widths
given in Ref.\cite{PDG} in order to emphasize the pseudoscalar meson exchange dominance in 
exciting the $S_{11}(1535)$ resonance.
 Since the $P_{11}(1440)$
resonance is below the $\eta N$ threshold, the corresponding coupling constant $g_{\eta NN^*}$
has been determined by folding the results with the mass distribution of the resonance which is
assumed to be given by a Breit-Wigner form. The signs of the coupling constants are chosen 
consistently with those used in the $\pi$ and $\eta$ photo-production analysis 
\cite{Benmer,RCOUP}.  
  
Following Ref.\cite{Nak3} and, in complete analogy to the nucleonic current, 
we introduce the off-shell form factors at each vertex involved in the resonance 
current. We adopt the same form factor given by Eq.(\ref{formfactorN}), with 
$m_N$ replaced by $m_{N^*}$ at the $MNN^*$ vertex, in order to account for the 
$N^*$ resonance being off-shell. The $MNN^*$ vertex, where the exchanged 
meson is also off-shell, is multiplied by an extra form factor $F_M(q^2)$ in 
order to account for the off-shellness of this meson 
(see Eqs.(\ref{transpot12},\ref{transpot32})). The corresponding full form factor 
is, therefore, given by the product $F_N(p^2)F_M(q^2)$, where $M$ stands for the 
exchanged meson between the two interacting nucleons. The form factor $F_M(q^2)$ 
is taken consistently with the $NN$ potential $V$ in Eq.(\ref{scateqn}); the only 
two differences are the normalization point of $F_v(q^2) (v=\rho,\omega)$ and
the cutoff parameter value of $F_\pi(q^2)$. Here, the form factor for vector mesons  
$F_v(q^2)$ is normalized to unity at $q^2=0$  in accordance with the kinematics at 
which the coupling constant $g_{vNN^*}$ was extracted, i.e., 
$F_v(q^2) = (\Lambda_v^2 / (\Lambda_v^2 - q^2))^2$. For the pion form factor 
$F_\pi(q^2)$, following the discussion in 
Refs.\cite{Nak1,Nak2,Nak3}, we use the cutoff value of $\Lambda_\pi=0.9 ~GeV$. We 
also use this value of the cutoff in the form factor at the $\pi NN$ vertex in 
Eqs.(\ref{transpot12},\ref{transpot32}) as well as in the $\pi NN$ vertex appearing
in the mesonic current constructed in the next subsection.

\subsection{The Mesonic Current}

For the meson-exchange current we consider the contribution from the 
$\eta vv$ vertex with $v$ denoting either a $\rho $ or $\omega $ meson. 
This gives rise to the dominant meson-exchange current. 
The $\eta vv$ vertex required for constructing the meson-exchange current 
is derived from the Lagrangian densities
\begin{eqnarray}
{\cal L}_{\eta\rho\rho}(x) & = & - \frac{g_{\eta\rho\rho}}{2 m_\rho}
\varepsilon_{\alpha\beta\nu\mu} (\partial^\alpha \vec \rho^\beta(x)) \cdot
(\partial^\nu \vec \rho^\mu(x)) \eta(x) \nonumber \\
{\cal L}_{\eta\omega\omega}(x) & = & - \frac{g_{\eta\omega\omega}}
{2 m_\omega}\varepsilon_{\alpha\beta\nu\mu} (\partial^\alpha \omega^\beta(x)) 
(\partial^\nu \omega^\mu(x)) \eta(x)  \ ,
\label{vveta}
\end{eqnarray}
where $\varepsilon _{\alpha \beta \nu \mu }$ is the Levi-Civita antisymmetric 
tensor with $\varepsilon _{0123}=-1$. The vector meson-exchange current is then 
given by
\begin{equation}
J_{\eta vv} = \sum_{v=\rho ,\omega}\left\{[\Gamma^\alpha_{vNN}(k_v)]_1 
iD_{\alpha\beta}(k_v)
              \Gamma^{\beta\mu}_{\eta vv}(k_v, k^\prime_v)
              iD_{\mu\nu}(k^\prime_v) [\Gamma^\nu_{vNN}(k^\prime_v)]_2 
\right\} \ ,
\label{vveta_cur}
\end{equation}
where $D_{\alpha \beta }(k_{v })$ and $D_{\mu\nu}(k^\prime_v)$ stand for the (Feynman) 
propagators of the two exchanged vector mesons (either the $\rho $ or $\omega $ mesons as 
$v =\rho $ or $\omega $) with four-momentum $k_{v }$ and $k^\prime_v$, respectively. The 
vertices involved in the above equation are self-explanatory. The $vNN$ vertex
$\Gamma _{vNN}^{\mu}(v =\rho ,\omega )$ is taken consistently with those in the potential 
used for constructing the $NN$ T-matrix in Eq.(\ref{scateqn}). The $\omega NN$ coupling 
constant is, however, taken to have the same value mentioned in the previous subsection.

The coupling constant $g_{\eta vv}$ is determined from a systematic analysis of the
radiative decay of pseudoscalar and vector mesons in conjunction with VMD.
This is done following Refs.\cite{Nak2,Nak3,Durso}, with the aid of an 
effective Lagrangian with SU(3) flavor symmetry and imposition of the OZI rule \cite{OZI}. 
The parameters of this model are the angle $\alpha_V (\alpha _{P})$, which measures the 
deviation from the vector(pseudoscalar) ideal mixing angle, and the coupling constant of 
the effective SU(3) Lagrangian. They are determined from a fit to radiative decay of 
pseudoscalar and vector mesons. The parameter values determined in this way in 
Ref.\cite{Durso} (model B), however, overpredict the measured radiative decay width of 
the $\eta^\prime$ meson\cite{PDG}. Therefore, we have readjusted slightly the value of 
the coupling constant of the SU(3) Lagrangian in order to reproduce better the measured 
widths. We have $\alpha _{V}\cong 3.8{{}^{\circ }}$ and $\alpha _{P}\cong -45{{}^{\circ }}$, 
as given by the quadratic mass formula, and the coupling constant of the effective SU(3) 
Lagrangian of $G=7$ in units of $1/\sqrt{m_{v }m_{v }^{\prime }}$, where $m_{v }$ and 
$m_v^\prime$ stand for the mass of the two vector-mesons involved. The sign of the 
coupling constant $G$ is consistent with the sign of the $\rho\pi\gamma$ and 
$\omega\pi\gamma$ coupling constants taken from an analysis of the pion photoproduction 
data in the $\sim 1~GeV$ energy region \cite{Garc}. With these parameter values we obtain
\begin{eqnarray}
g_{\eta\rho\rho} & = & G \cos(\alpha_P) = 4.94 \nonumber \\
g_{\eta\omega\omega} & = & -G\left( \sqrt{2}\sin^2(\alpha_V)\sin(\alpha_P)
+ \cos^2(\alpha_V)\cos(\alpha_P) \right) = 4.84 \ .
\label{meccoupl}
\end{eqnarray}

The $\eta vv$ vertex $(v =\rho ,\omega )$ in Eq.(\ref{vveta_cur}), where the 
exchanged vector mesons are both off their mass shells, is accompanied by a form factor. 
Following Ref.\cite{Nak2,Nak3}, we assume the form
\begin{equation}
F_{\eta vv}(k_v^2, {k^\prime_v}^2) = 
\left( \frac{\Lambda_v^2 - m_v^2} {\Lambda_v^2 - k_v^2} \right)
\left ( \frac{\Lambda_v^2} {\Lambda_v^2 - {k^\prime_v}^2} \right ) \ .
\label{formfactorM}
\end{equation}
It is normalized to unity at $k_{v }^{2}=m_{v }^{2}$ and $k_{v }^{\prime \text{ }2}=0$, 
consistent with the kinematics at which the value of the coupling constant 
$g_{\eta vv}$ was determined. We adopt the cutoff parameter value of $\Lambda _{v}
= 1.45 ~GeV$ as determined in Ref.\cite{Nak2} from the study of the $\omega$ and 
$\phi$ meson production in $pp$ collisions. This form factor has been also used in the 
study of the $\eta^\prime$ meson production in Ref.\cite{Nak3}.

Another potential candidate for mesonic current is the $\eta a_o\pi$-exchange
current, whose coupling constant may be estimated from the decay width of $a_o$
into an $\eta $ and $\pi$. We take the Lagrangian density
\begin{equation}
{\cal L}_{\eta a_o\pi}(x)  =  
\frac {g_{\eta a_o\pi}} {\sqrt{m_{a_o} m_\pi}}
\vec a_o(x)\cdot (\partial_\mu\vec\pi(x))(\partial^\mu\eta(x)) \ ,
\label{aopsps}
\end{equation}
where $\vec a_o(x)$ stand for the $a_o$ meson field. $m_\pi$ and $m_{a_o}$
stand for the masses of the $\pi$ and $a_o$ meson, respectively.  
Using the measured decay width from \cite{PDG} we obtain a value of 
$\left| g_{\eta a_o\pi} \right| \simeq 1.81$. The sign of this coupling constant is 
not fixed. We assume it to be positive in the present work. Since the contribution of
the $\eta a_o\pi$ current is small, the sign of its coupling constant will not affect
the major conclusion of the present work.

The $\eta a_o\pi$ current reads
\begin{equation}
J_{\eta a_o\pi} = \left\{[\Gamma_{a_oNN}]_1
i\Delta_{a_o}(k_{a_o}^2)
              \Gamma_{\eta a_o\pi}(k_{a_o} , k_\pi)
              i\Delta_\pi(k_\pi^2) [\Gamma_{\pi NN}(k_\pi)]_2 
\right\} +  (1 \leftrightarrow 2) \ 
\label{etaaopi_cur}
\end{equation}
in our previously-defined notation. The vertex $\Gamma_{a_oNN}$ is taken consistently 
with that in the $NN$ potential $V$ in Eq.(\ref{scateqn}), while for the $\pi NN$ 
vertex we use the same one (including the cutoff parameter value) mentioned in the 
previous subsection. The $\eta a_o\pi$ vertex, $\Gamma_{\eta a_o\pi}(k_{a_o} , k_\pi)$, 
is assumed to have a form factor given by 
\begin{equation}
F_{\eta a_o\pi}(k_{a_o}^2, {k_\pi}^2) = 
\left( \frac{\Lambda_{a_o}^2 - m_{a_o}^2} {\Lambda_{a_o}^2 - k_{a_o}^2} \right)
\left ( \frac{\Lambda_\pi^2 - m_\pi^2} {\Lambda_\pi^2 - {k_\pi}^2} \right )
\label{formfactorM1}
\end{equation}
with $\Lambda_{a_o} = \Lambda_\pi = 1.45 \ GeV$.

The total mesonic current is then given by
\begin{equation}
J_{mec} = \sum_{v=\rho,\omega}J_{\eta vv} + J_{\eta a_o\pi} \ .
\label{mec_cur}
\end{equation}

There are, of course, other possible mesonic currents, such as the $\eta \omega
\phi$- and $\eta \phi \phi$-exchange currents, that contribute to $\eta$ meson 
production in $NN$ collisions. Their contributions have been estimated in a systematic 
way following Ref.\cite{Nak2} and were found to be negligible.

\vfill \eject

\begin{table}
\caption{Coupling constants $g_{MNN^*}$ used in the resonance current.
Those in parenthesis for vector mesons refer to 
($g_{MNN^*}^{(1)}, g_{MNN^*}^{(2)}$). The ps-pv mixing parameter is 
fixed to be $\lambda = 0$ and the off-shell parameter $z=-1/2$. 
The masses and widths of the resonances, $m_{N^*}$ and $\Gamma_{N^*}$,
are in units of $MeV$.}
\vskip -0.25cm
\tabskip=1em plus2em minus.5em
\halign to \hsize{\hfil\bf#&
                  \hfil#\hfil&\hfil#\hfil&\hfil#\hfil\cr
\noalign{\hrulefill}
$N^*$                 & $S_{11}(1535)$ & $P_{11}(1440)$ & $D_{13}(1520)$ \cr
$(m_{N*},\Gamma_{N^*})$ & (1535,150)   &  (1440,350)    &    (1520,120)  \cr
\noalign{\vskip -0.20cm}
\noalign{\hrulefill}
\noalign{\vskip -0.5cm}
\noalign{\hrulefill}
$\pi$                &  1.25 &  6.54 &  1.55 \cr
$\eta$               &  2.02 &  0.49 &  6.30 \cr
$\rho$               & -0.65 & -0.57 & (6.0, -2.1)  \cr
$\omega$             & -0.72 & -0.37 & (-2.1, 0.7) \cr
\noalign{\hrulefill} }
\label{tab1}
\end{table}

\vglue 0.5cm

\newpage
\vglue 0.5cm

\newpage
\vglue 0.5cm

\newpage
\vglue 0.5cm

\newpage
\vglue 0.5cm

\newpage
\vglue 0.5cm

\newpage
\vglue 0.5cm

\newpage
\vglue 0.5cm


\begin{thebibliography}{99}
\bibitem{notelee}
We note that if the full two-nucleon propagator is 
used, the amplitude in Eq.(\ref{ampl0}) would have an additional term to avoid the 
double counting which otherwise would arise from the terms involving 
$i G^{(+)}_i T^{(+)}_i$ and/or $T^{(-)\dagger}_f i G^{(-)*}_f$ and
the nucleonic current $J_{nuc}$ as given in Eq.(\ref{curr}) and constructed explicitly 
in the appendix (see also Fig.~\ref{fig1}). This additional term vanishes if the full
two-nucleon propagator is approximated by a three dimensionally reduced propagator
$G_{(i/f)}$ which restricts the energies of the propagating two nucleons onto their 
respective mass-shell energies.
\end{thebibliography}
\end{document}